\documentclass[sigconf,authorversion,nonacm]{acmart}
\usepackage{pdflscape}
\usepackage{multirow}
\usepackage{geometry}

\usepackage{float}

\usepackage [english]{babel}
\usepackage [autostyle, english = american]{csquotes}
\MakeOuterQuote{"}

\usepackage{subcaption}
\usepackage{graphicx}
\usepackage{tikz}
\usepackage{amsmath}
\usepackage{float}

\AtBeginDocument{%
 }

\begin{document}

\title{The Veracity Problem: Detecting False Information and its Propagation on Online Social Media Networks}

\author{Sarah Condran}
\email{scondran@csu.edu.au}
\orcid{0000-0001-7813-2116}
\affiliation{%
  \institution{School of Computing, Mathematics and Engineering, Charles Sturt University}
  \city{Wagga Wagga}
  \state{NSW}
  \country{Australia}
  \postcode{2650}
}

\renewcommand{\shortauthors}{Sarah Condran}

\begin{abstract}
Detecting false information on social media is critical in mitigating its negative societal impacts. To reduce the propagation of false information, automated detection provide scalable, unbiased, and cost-effective methods. However, there are three potential research areas identified which once solved improve detection. First, current AI-based solutions often provide a uni-dimensional analysis on a complex, multi-dimensional issue, with solutions differing based on the features used. Furthermore, these methods do not account for the temporal and dynamic changes observed within the document's life cycle. Second, there has been little research on the detection of coordinated information campaigns and in understanding the intent of the actors and the campaign. Thirdly, there is a lack of consideration of cross-platform analysis, with existing datasets focusing on a single platform, such as X, and detection models designed for specific platform.  

This work aims to develop methods for effective detection of false information and its propagation. To this end, firstly we aim to propose the creation of an ensemble multi-faceted framework that leverages multiple aspects of false information. Secondly, we propose a method to identify actors and their intent when working in coordination to manipulate a narrative. Thirdly, we aim to analyse the impact of cross-platform interactions on the propagation of false information via the creation of a new dataset.  
\vspace{-0.5em}
\end{abstract}


\maketitle

\begin{figure*}[th]
    \centering
    \includegraphics[scale=0.057]{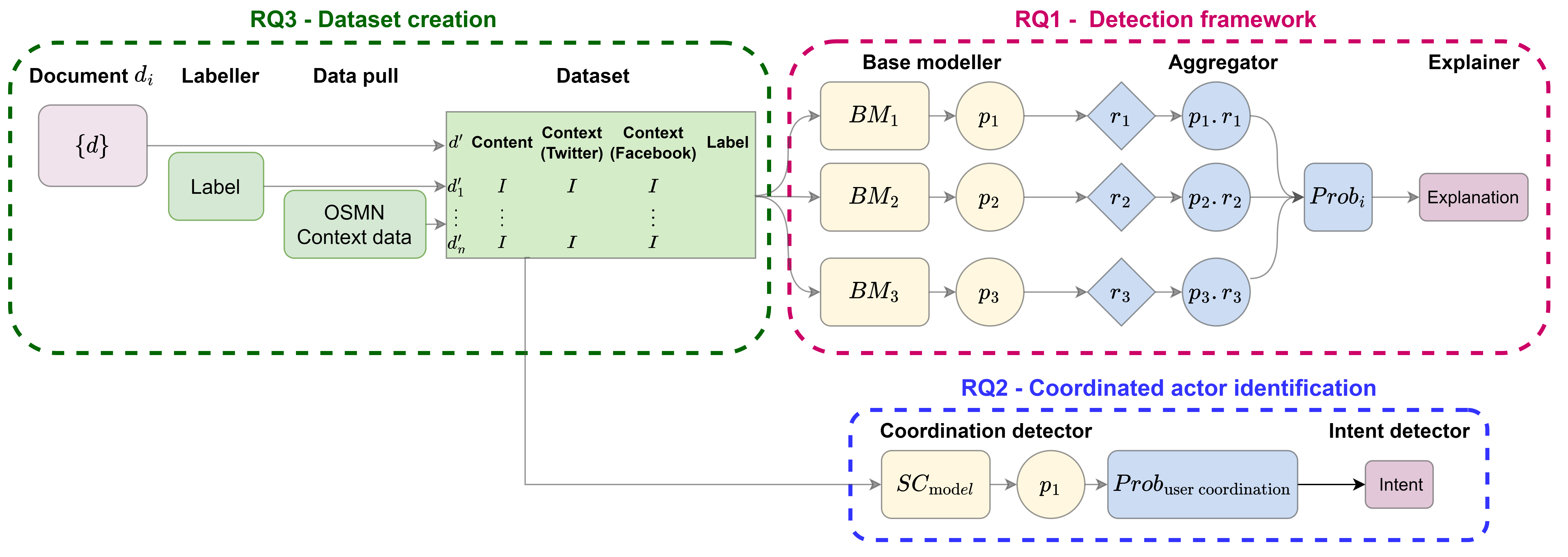}
    \caption{The Proposed Methodology}
    \label{figureMethodology}
    \vspace{-1em}
\end{figure*}

\section{Introduction}
The creation and spread of false information (\emph{aka} fake news) is rapidly increasing, with online social media networks (OSMN) such as X\footnote{Formerly known as Twitter}, Facebook, and Weibo contributing to its rise. False information can, often unbeknownst to them, manipulate how individuals responds to topics such as health, politics, and social life. One such example is a tweet in 2013 that claimed an explosion injured the US president, resulting in a loss of \$130 billion in stock values \cite{Elboghdardy2013}. Further, some research suggests false information is connected to election outcomes e.g. the 2016 US presidential elections \cite{WTOE2016} and 2020 UK Brexit vote \cite{Assenmacher2020}.

OSMN enables anyone to access the latest information in a variety of formats (\emph{i.e.} news articles, blogs) and sources (\emph{i.e.} news outlets, public figures). The expansion of new formats and sources has decentralised the distribution of information, removing centralised control over its authenticity. This unregulated creation and spread of information places the onus of validating the truthfulness (or falsehood) of the information on the individual.
However, an individual's ability to identify falsehoods objectively is influenced by factors such as \textit{confirmation bias}, which makes one trust and accept information that confirms their preexisting beliefs \cite{Nickerson1998}, and \textit{selective exposure}, which is when one prefers to consume information that aligns with their beliefs \cite{Freedman1965}. External factors such as the \textit{bandwagon effect}, which motivates one to perform an action because others are doing it \cite{Leibenstein1950}, and the \textit{validity effect}, where one believes information after repeated exposure \cite{Boehm1994} play a critical role. 

A recent narrative which exemplified coordinated behaviors and the \emph{bandwagon effect} is the Sydney Bondi Junction Stabbings \cite{Nguyen2024_BondiStabbings}. An image of the attacker circulated on social media, resulting in widespread misidentification of his heritage, religion, and identity. This false information initiated from small accounts, was amplified by verified accounts, and reported on by mainstream news. The coordinated behaviour exhibited both malicious and benign propagation patterns.

\section{Literature Review}\label{ch:literature_review}
Detecting false information is crucial in alleviating the burden placed on individuals to verify the truthfulness of information, and the resulting consequences of this burden. There are three main streams of research key to this work which aims to address the problem of false information. These relate to (1) false information detection methods; (2) detection of coordinated behaviours; (3) cross-platform analysis of false information propagation. 

\textbf{False Information Detection Methods}:
To overcome the limitations of \emph{human-driven} false information detection (\emph{e.g.} scalability and bias \cite{Tchakounte2020, Wallace2022_DLA}), AI-based decision support techniques have been developed. Most techniques fall into two types: content-based methods and context-based methods. Content-based methods are built using the information contained within the document, such as words and images \cite{Zhang2024_HeteroSGT, Ghanem2021_FakeFlow, Shu2019_dEFEND, Wang2024_EFND, Jin2022_FinerFact}. While these methods can preemptively forestall the dissemination of false information, they are prone to adversarial manipulation of linguistic and stylistic features to evade detection \cite{Jin2022_FinerFact}. On the other hand, context-based methods use the information on the document's propagation over OSMNs and the users who engage with it \cite{Soga2024_stance, Chowdhury2020_CSM, Bing2022, Min2022_PSIN, Ruchansky2017_CSI}. These methods are independent of linguistic and stylistic features and knowledge bases, but rely on information generated when a document is propagated on an OSMN, meaning false information is already spread, along with its related negative consequences. Although some hybrid models such as \emph{dEFEND}~\cite{Shu2019_dEFEND} and \emph{CSI}~\cite{Ruchansky2017_CSI} have been proposed, they use limited sources of information, making them prone to loss of reliability and effectiveness.

\textbf{Detecting Coordination}:
There are two approaches to detect coordination, the first considers the similarity of actors \cite{Assenmacher2020, Nizzoli2022}. For example, \cite{Nizzoli2022} proposed a network-based framework using iterative community detection to estimate the extent of coordination among actors. The second method considers the temporal synchronisation of actors. For example, \cite{Zhang2021_vigdet} incorporated a neural point process to identify synchronised behaviours. Moreover, most detection works treat the threshold of coordination as a distinct boundary within detection methods. However, it is a spectrum and varies for each community of actors \cite{Giglietto2020}. Methods that attempt to distinguish intent have found an inability to differentiate \cite{Pacheco2021}. While \cite{Fazil2020} assumed all coordinated behaviours indicated inauthenticity, no other works were found that identify the intent of an actors when engaging in coordinated behaviours.

\textbf{Existing Datasets}:
The available datasets for developing detection methods vary largely in terms of features for each document. For example, \textit{FakeNewsNet} \cite{Shu2020_FakeHealthNetDataset} provides both context and content data, while \textit{LIAR} provides only content data. This means models developed on one dataset are often not transferable to other datasets. Additionally, most datasets include context data from one OSMN, making cross-platform analysis challenging. However \cite{Ma2016_Ma} does include context data from both X and Weibo.
The labelled datasets use manual labelling of documents and often require the manual identification of items corresponding to the document. Consequently, these datasets are static, and often out of date, leading to domain drift and diminished model performance in new domains. 

\section{Research Aim, Gaps and Questions}
The aim of this research is to develop methods for the effective detection of false information on social media, and the analysis thereof, of its propagation.

The first research gap identified pertains to the detection of false information propagated on OSMNs. The existing methods consider the context and content features in isolation which provides a uni-dimensional view on a multi-dimensional concept. For example, in one dimension (e.g. content) an document may be labelled as true while in another (r.g. context), that same document may be labelled false. Further, the available data for a document and performance of a model can be impacted by the temporal and dynamic aspect of real world data. For example, consider a document observed at two times ($d_1.t_1$, $d_1.t_{100}$) where $d_1.t_1$ has 3 engagements and $d_1.t_{100}$ has 500 engagements. A model based on context data (\emph{e.g.} number of user engagements) may assign a falsehood probability to $d_1.t_1$ based on significantly fewer data points compared to $d_1.t_{100}$, thus affecting the quality of the prediction. This motivates the research question: \textbf{RQ1} \emph{How can a model agnostic framework be developed to improve explainability and accuracy of existing false information detection techniques without incurring excessive computational overheads?}

The second research gap identified is the detection of coordinated behaviours. The few works that detect coordination campaigns or actors are built on static datasets covering a limited range of topics and campaigns. However, no works look at identifying coordination on streaming data or without predefined campaigns to search for. Furthermore, no work has distinguished the intent of coordinated behaviours. This motivates the research question: \textbf{RQ2} \emph{How can we identify actors working in coordination to maliciously amplify and manipulate the creation and propagation of information?}

The third research gap is a lack of consideration of cross-platform interactions in detecting false information. There are no works to date which merge the context data from multiple OSMNs with the content data for each document to enhance the data available for analysis. Further, few works analyse whether the existing detection models are suitable for the different OSMNs. This motivates the research question: \textbf{RQ3} \emph{ How might the propagation of false information change over different online social media networks, and can existing detection methods account for the differences?}

\section{Methodology}
The overarching aim of this work is addressed through the proposed methodology summarised in Figure \ref{figureMethodology}, which outlines the approaches to solve the three research questions. The first research question \textit{RQ1 - Detection framework)}, will be solved through a three-part framework. This framework takes a document and its associated features as input, producing a probability of falsehood and an explanation as the output. (1) \textbf{Base modeller}: This component employees various false information detection base models (BMs) to generate probabilities of falsehood $(p)$. (2) \textbf{Aggregator}: This component combines the various probabilities $(p)$ into a final falsehood probability $(Prob)$. A novel dynamic aggregation method will be developed which accounts for the reliability of features the prediction is based on. For each instance of a document, a reliability weight $(r)$ will be assigned the the probability $(p)$. (3) \textbf{Explainer}: This component provides a tiered explanation of the contributions to the final probability $(Prob)$. Each tier offers greater detail of the BMs and the reliability factors.

The second research question \textit{RQ2 - Coordinated}, solution is a two-part detection method which will first identify actors working in coordination and then determines the intent behind their behaviours. (1) \textbf{Coordinated detector}: This component identify the actors contributing to the a coordination campaign using isolated user characteristics (\emph{e.g.} URLs, user mentions, hashtags) rather then explicit network structures (\emph{e.g.} follower network, retweet information) which are often incomplete or unavailable. The pre-training and fine-tuning of models to reduce the required training dataset will be incorporated. (2) \textbf{Intent detector}: This component employs contrastive learning to identify an actors intent. That is, whether the actors harmful intent with malicious behaviours, or benign where an actor is engaging in coordinated activities without maliciousness for example as part of the \emph{bandwagon effect}.
 
The third research question \textit{RQ3 - Dataset creation}, solution is a three-part process to incorporate a new document into a multi-OSMN dataset. The dataset once created can enable the analysis of cross platform interactions. (1) \textbf{Labeller}: This segment uses an API to obtain a label for each document from a variety of manual factchecking sites. (2) \textbf{Data Pull}: This segment uses an API which collects all related context data from various OSMNs for each document. This novel method will enable the identification of items on a OSMN without prior manual identification. (3) \textbf{Dataset}: This segment will merge multiple context features and content features for one comprehensive data package for each document. Further, this segment employs entity linking to merge a user's behaviour from multiple OMSNs. 

\section{Preliminary Results}
Preliminary experimentation for the development of the \textbf{Aggregator} (RQ1) have returned favourable results for the use of a dynamic and adaptive weighting scheme. A brief outline of the experimental setup and results are presented below.

\noindent\textbf{Datasets}: Consistent with existing works, the benchmark datasets \emph{PolitiFact} \cite{Shu2020_FakeHealthNetDataset}, \emph{GossipCop} \cite{Shu2020_FakeHealthNetDataset}, and \emph{FakeHealth} \cite{Dai2020_FakeHealthDataset} are used to develop this aggregation method.

\noindent\textbf{Base Models(BM)}: Three state-of-the-art base models are considered: (1) \textit{Fakeflow} (FF) \cite{Ghanem2021_FakeFlow} utilises content features with Bidirectional Gated Recurrent Units (Bi-GRUs) to learn the flow of affective information throughout a document; (2) \emph{Publisher Credibility} (PC) adapted from \cite{Chowdhury2020_CSM} utilises document publisher features to train a probabilistic soft logic (PSL) model to calculate the credibility of a publisher; and (3) \emph{User Credibility} (UC) adapted from \cite{Chowdhury2020_CSM} utilises context features in the form of users who create an item to train a PSL model to calculate the credibility of that user. 

\noindent\textbf{Setup}: The framework was built using python 3 running on Linux. And for all BMs the train-validation-test split was 70-10-20, with 10-fold cross validation.

\begin{figure}
    \centering
    \includegraphics[scale=0.4]{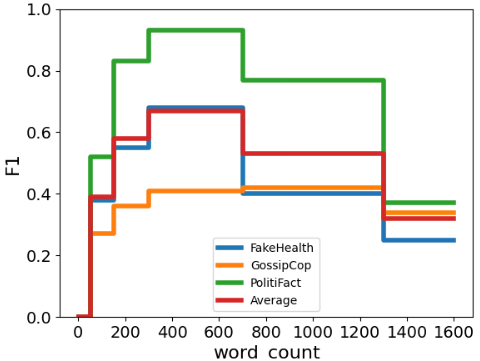}
    \caption{Reliability factor \emph{word\_count}}
    \label{figure_F1_wordcount}
    \vspace{-1.65em}
\end{figure}

\noindent\textbf{Aggregator}: 
To develop a dynamic aggregation method based on the reliability of features, a set of reliability factors are produced for document $d_i$. The factors are defined on how they influence the features which contributes to a models prediction of falsehood. For instance, for context-based models such as \emph{FF}, the features required are a set of words for each document $d_i$. Thus we make the assumption that a document $d_i$ with a reasonable amount of words will provide \emph{FF} with "sufficient" information to make a prediction of falsehood. This intuition is supported by the experiments shown in Figure \ref{figure_F1_wordcount} as measured by $F1$ scores for three datasets and different text lengths using \emph{FF}.

\begin{figure*}[!t]
    \centering
    \captionsetup{skip=0pt} 
    \includegraphics[scale=0.058]{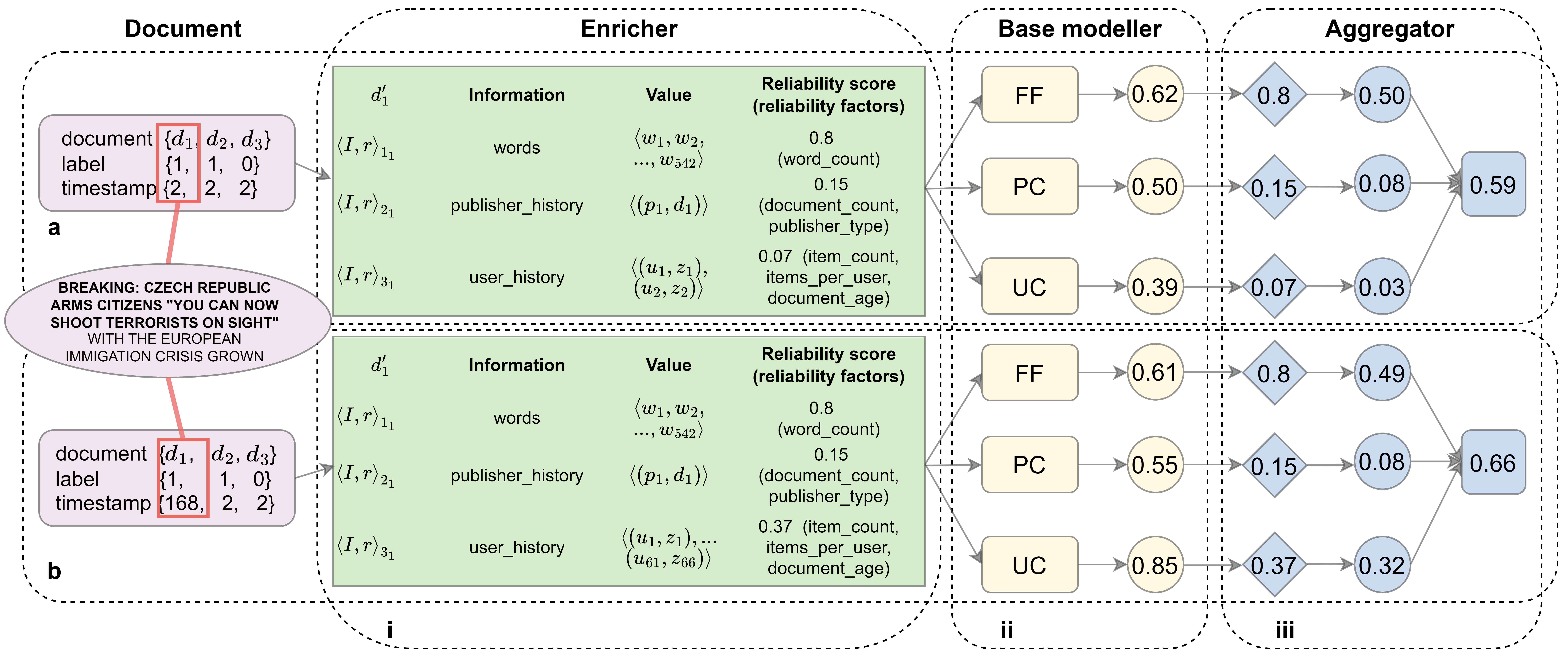}
    \caption{The Detection Framework}
    \label{figureFramework}
\end{figure*}

\noindent\textbf{Results}: 
A detailed example (Figure \ref{figureFramework}) shows the working of the dynamic aggregation method where the weights are adaptively calculated for each instance of the document. Specifically, \ref{figureFramework}.a.iii illustrates the dynamic weightings assigned to the three base models for document $d_1$ at time $t_2$. In Figure \ref{figureFramework}.b.iii the same document $d_i$ at $t_{168}$, due to a higher level of user engagements due to the age of the document, has different weightings assigned to models which use \emph{user\_history} features. That is, \textit{UC} had a higher contribution to the final prediction of falsehood.

\section{Conclusion}
In this work, we identify three critical research gaps and propose solutions in the detection and mitigation of false information on social media. The proposed framework, once implemented, will improve false information detection by leveraging multiple base models in a dynamic manner. This will include developing novel methods for dynamic data aggregation based on reliability and providing hierarchical tiered explanation. The second proposed method involves a novel algorithm to identify both coordinated actors and their intent. Finally, the proposed dataset aims to provide a comprehensive view of a document and enable interrogation of cross-platform interactions. 

Future research directions for the first research question include exploring additional reliability factors across a broader range of base models to determine whether there is a peak or a plateau in performance as reliability factors change. Another direction is to test the proposed framework on different online social media networks to assess its versatility and determine whether it is truly model-agnostic.

\section{Acknowledgements}
I would like to thank my supervisors Dr Michael Bewong, Dr Selasi Kwashie, Professor Md Zahidul Islam, and Associate Professor Irfan Altas for their support and guidance. 
This work was supported by Charles Sturt University.

\balance

\bibliographystyle{ACM-Reference-Format}

\bibliography{Proposal/arxiv/paper}

\end{document}